\documentclass[a4paper]{jpconf}
\usepackage{graphicx}

\usepackage[utf8]{inputenc}          
\usepackage{graphicx,xcolor}         
\usepackage{array,dcolumn,longtable} 
\usepackage{amsmath,amssymb,slashed} 

\providecommand\lfstyle{}                   
\providecommand\romanup[1]{\text{#1}}       
\renewcommand\textsc{\MakeUppercase}

\usepackage{siunitx}                                  
\DeclareSIUnit{\fm}{\femto\metre}                     

\newcommand\dabmod{\textsc{dab-mod}}
\newcommand\vusphydro{\text{v-\textsc{usp}hydro}}

\newcommand\qcd{\textsc{qcd}}

\newcommand\fonll{\textsc{fonll}}

\declareslashed{}{\divslash}{0.08}{0}{\mathcal{D}}

\DeclareSIUnit{\fm}{\femto\metre}

\newcommand\PbPb{{\romanup{PbPb}}}

\newcommand\ArAr{{\romanup{ArAr}}}
\newcommand\XeXe{{\romanup{XeXe}}}
\newcommand\OO{{\romanup{OO}}}

\newcommand\Xe{{\romanup{Xe}}}

\newcommand\snn[1][]{\sqrt{s_\text{NN}}\ifx\\#1\\\else=\SI{#1}{\TeV}\fi}

\newcommand\raa{R_\text{AA}}

\newcommand\Tfo{T_\text{FO}}

\begin{document}
\title{Heavy flavor dynamics across system size at the LHC}

\author{Roland Katz}
\address{SUBATECH, Universit\'e de Nantes, EMN, IN2P3/CNRS, 44307 Nantes, France}
\author{Caio A.~G.~Prado}
\address{Institute of Particle Physics, Central China Normal University (CCNU), Wuhan, Hubei 430079, China}
\author{Jacquelyn Noronha-Hostler}
\address{University of Illinois at Urbana-Champaign, Urbana, IL 61801, USA}
\ead{jnorhos@illinois.edu}
\author{Alexandre A.~P.~Suaide}
\address{Instituto de F\'{i}sica, Universidade de S\~{a}o Paulo, C.P. 66318, 05315-970 S\~{a}o Paulo, SP, Brazil}

\begin{abstract}
One of the fundamental signatures of the Quark Gluon Plasma has been the suppression of heavy flavor (specifically D mesons), which has been measured via the nuclear modification factor, $R_{AA}$ and azimuthal anisotropies, $v_n$, in large systems. However, multiple competing models can reproduce the same data for $R_{AA}$ to $v_n$.  In this talk we break down the competing effects that conspire together to successfully reproduce  $R_{AA}$ and $v_n$ in experimental data using Trento+v-USPhydro+DAB-MOD. Then using our best fit model we make predictions for $R_{AA}$ and $v_n$ across system size for $^{208}\PbPb$, $^{129}\XeXe$, $^{40}\ArAr$, and $^{16}\OO$ collisions. We find that 0--10\% centrality has a non-trivial interplay between the system size and eccentricities such that system size effects are masked in $v_2$ whereas in 30--50\% centrality the eccentricities are approximately constant across system size and, therefore, is a better centrality class to study  D meson dynamics across system size.
\end{abstract}

\section{Introduction}

The suppression of heavy probes has been a fundamental signature used to constrain the microscopic properties of the Quark Gluon Plasma (QGP). The soft sector of the QGP consists of a nearly perfect fluid composed of light quarks (up, down, and strange) and gluons. The collective flow observables in the soft sector are well described through event-by-event relativistic viscous hydrodynamics \cite{Song:2010mg, Bozek:2012qs, Gardim:2012yp, Bozek:2013uha, Niemi:2015qia, Ryu:2015vwa, McDonald:2016vlt, Bernhard:2016tnd, Gardim:2016nrr, Giacalone:2016afq, Alba:2017hhe, Gardim:2017ruc,Giacalone:2017dud, Eskola:2017bup, Weller:2017tsr, Schenke:2019ruo,Giacalone:2020lbm,Giacalone:2020dln}. Due to the much larger mass of charm quarks, they are created at very short time scales and are very unlikely to be fully thermalized within the soft sector of the QGP.  Therefore, open heavy flavor mesons, like D mesons can provide orthogonal information to that gathered in the soft sector.  Studying the nuclear modification factor, $R_{AA}$, which is approximately 1 when no suppression occurs and much less than 1 in large AA collisions where heavy flavor quarks loose energy by being bumped around by the strongly interacting QGP, can provide insight into the microscopic dynamics of the QGP.  Additionally, due to the path length dependence of the heavy quarks passing through the azimuthally anisotropic medium, a large $v_2$ is measured for D mesons. In more recent years it has been shown that there is a strong correlation between the soft and hard/heavy sectors when it comes to azimuthal anisotropies \cite{Nahrgang:2014vza,betz:2016ayq,noronha-hostler:2016eow, prado:2016szr, sirunyan:2017pan,Andres:2019eus,Katz:2019qwv}

In light of the recent small system results from the LHC and RHIC wherein collective flow and strangeness enhancement \cite{Chatrchyan:2013nka,Aaboud:2017acw,Aaboud:2017blb,Aad:2013fja,sirunyan:2018toe,Khachatryan:2014jra,Khachatryan:2015waa,Khachatryan:2015oea,Sirunyan:2017uyl,ABELEV:2013wsa,Abelev:2014mda,Adare:2013piz, Adare:2014keg,Aidala:2018mcw,Adare:2018toe,Adare:2015ctn,Aidala:2016vgl,Adare:2017wlc,Adare:2017rdq,Aidala:2017pup,Aidala:2017ajz,ALICE:2017jyt} have been measured in small systems that can be reasonably well described by theoretical hydrodynamic models \cite{Bozek:2011if,Bozek:2012gr,Bozek:2013ska,Bozek:2013uha,Kozlov:2014fqa,Zhou:2015iba,Zhao:2017rgg,Mantysaari:2017cni,Weller:2017tsr,Zhao:2017rgg}, physicists have been trying to reconcile these results with the lack of hard or heavy flavor suppression (i.e. $R_{pA}\sim1$ \cite{Adam:2015qda,Acharya:2019mno}). For an overview and discussion on future measurements see \cite{Perepelitsa:2020pcf}. A further piece in the puzzle is that a large $v_2$ has now been measured for D mesons in pPb collisions at the LHC \cite{Zhang:2019dth}. Currently, there are no theoretical models that can simultaneously reproduce $R_{pA}\sim1$ with a large $v_2$ prediction, rather $R_{pA}\sim 0.8$ \cite{Xu:2015iha} or no $v_2$ calculations have been attempted  \cite{Kang:2014hha,Sharma:2009hn}.  In order to shed light on this mystery we study the system size dependence of $R_{AA}$ and $v_2$ in the potential upcoming system size run of symmetric AA collisions (first proposed at the LHC in \cite{Citron:2018lsq}) using our best fit models in PbPb collisions: trento+v-USPhydro+DAB-MOD. 

\section{Model}

For our heavy flavor description we have created the modular D and B meson Monte Carlo simulation package \dabmod~\cite{prado:2016szr,Katz:2019fkc,Katz:2019qwv} that can run both Langevin and energy loss models (and easily incorporate different components of the modeling such as energy loss fluctuations, coalescence, varying the diffusion transport coefficient, and path length dependence of the energy loss).  The initial heavy flavor quarks are sampled from  p\qcd\ \fonll\ calculations~\cite{Cacciari:1998it,Cacciari:2001td} distribution. After they are evolved dynamically on top of an event-by-event viscous hydrodynamical evolving medium (Trento\,\cite{Moreland:2014oya}+\vusphydro\,\cite{Noronha-Hostler:2014dqa,Noronha-Hostler:2013gga,Noronha-Hostler:2015coa}) using either a Langevin description or energy loss, they are hadronized at the decoupling temperature $T_d$
via a hybrid fragmentation/coalescence model from which the final nuclear modification factor can be reconstructed. Here we use our ``best fit" model tuned at PbPb 5.02 TeV collisions \cite{Katz:2019fkc} where the purely collisional spatial diffusion coefficient model \cite{Moore:2004tg} $D_s(2\pi T)=2.23$ reasonably reproduces experimental data at low $p_T\lesssim 5$--\SI{6}{\GeV}, while an energy loss model (with the same best fit description as in \cite{Aaboud:2018bdg}) works best for  for the high $p_T\gtrsim 5$--\SI{6}{\GeV}. While a number of lessons can be learned from PbPb collisions, it is not clear across system size in what $p_T$ range Langevin versus energy loss descriptions will dominate.  Therefore, in the plots shown here we include both in an overlapping region.  

Because we also plot the azimuthal anisotropies, we point out that this is inherently a soft hard correlation (between one soft particle and one hard particle, see \cite{betz:2016ayq,noronha-hostler:2016eow, prado:2016szr, sirunyan:2017pan,Andres:2019eus} for further details on the implications). Therefore, we pay special attention to a reasonable hydrodynamic description of the soft sector.  Here we used Trento initial conditions with the parameters $p = 0$, $k = 1.6$, and $\sigma = \SI{0.51}{\fm}$ determined from a Bayesian analysis~\cite{Bernhard:2016tnd} and that works reasonably well compared to multiparticle cumulants \cite{Giacalone:2017uqx,Rao:2019vgy}.  The hydrodynamical description fixed the parameters $\tau_0 = \SI{0.6}{\fm}$, $\eta/s = 0.047$, $\Tfo = \SI{150}{\MeV}$ by comparing to experimental data from the LHC \cite{Alba:2017hhe,Giacalone:2017dud,Sievert:2019zjr} and uses the state-of-the-art Lattice QCD equation of state \cite{Borsanyi:2013bia,Alba:2017hhe} matched to the particle data list PDG16+ \cite{alba:2017mqu,Alba:2020jir}.  Additionally, due to the sensitivity of the soft sector to a deformed  $^{129}\Xe$ nucleus \cite{Giacalone:2017dud} we also compare a spherical versus prolate xenon nucleus.  

Finally, we would be remiss not to point out that there is a tuning parameter for  $\raa$ that adjusts its overall magnitude (generally tuned in most central collisions).  Because these are predictions for XeXe, ArAr, and OO i.e. there is no experimental data available for $\raa$ in these collisional systems, there is inherently an amount of uncertainty due to this parameter.  Furthermore, our hydrodynamic calculations have a similar parameter that is typically tuned to the $dN/dy$ in central collisions.  Here we assume in both cases that these constants do not vary with system size. 

\section{Results}

Within the heavy flavor sector there are many competing models that can reasonable well reproduce experimental data (see for these comparisons in \cite{Acharya:2017qps}). However, in \cite{Xu:2018gux} it was found that once all models are required to use the same hydrodynamical background that these similarities no longer hold.  In this work, we constrain our background to fit multiple observables in the soft sector and then study what contributes to produce a reasonable $\raa$ and $v_2$.

\begin{figure}[h]
\begin{minipage}{0.5\linewidth}
\includegraphics[width=\linewidth]{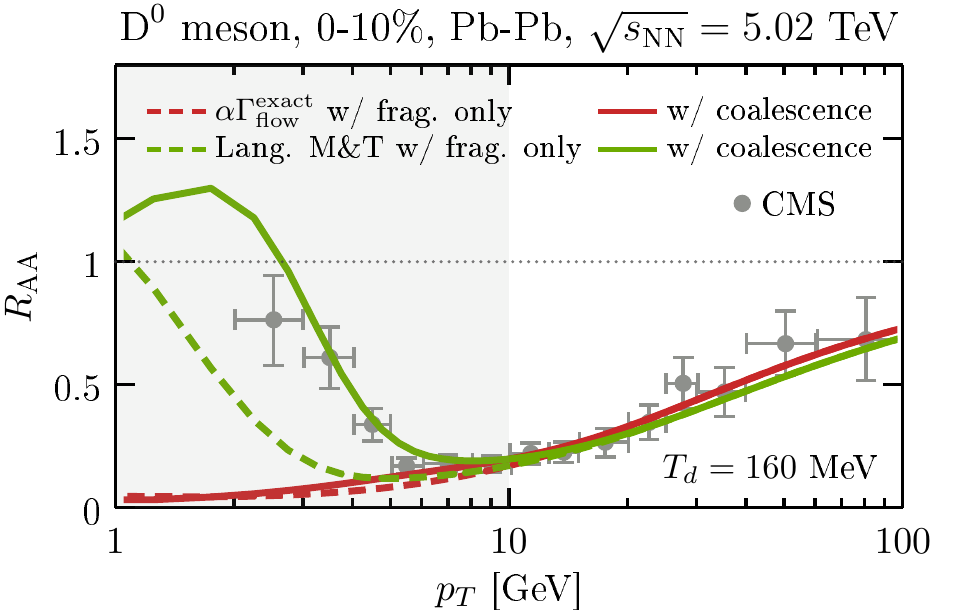}
\caption{\label{fig:raa_coal} Comparison of $R_{AA}$ for both our best fit Langevin and energy loss models both with and without coalescence.}
\end{minipage}\hspace{2pc}%
\begin{minipage}{0.5\linewidth}
\includegraphics[width=\linewidth]{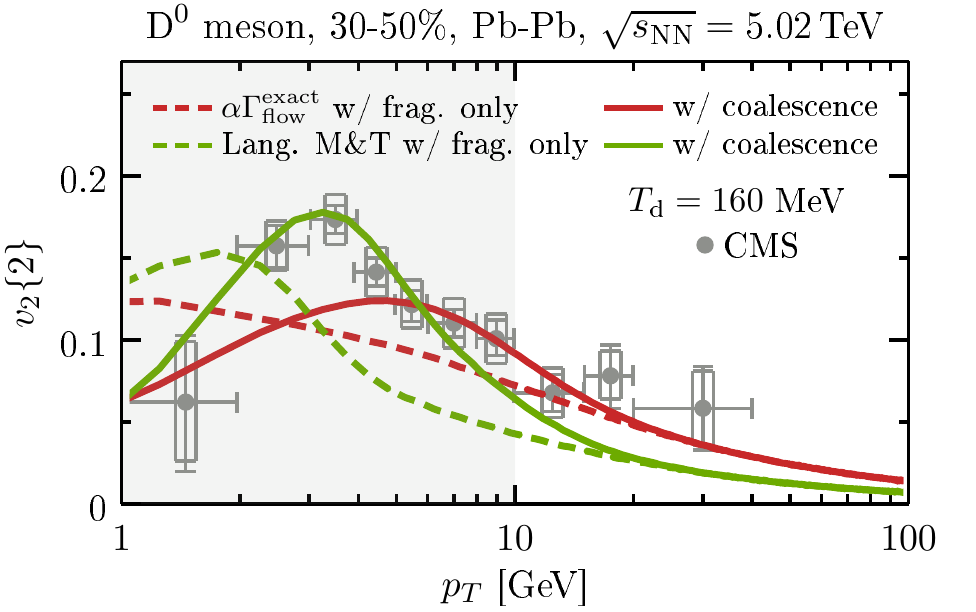}
\caption{\label{fig:v2_coal} Comparison of $v_2\left\{2\right\}$ for both our best fit Langevin and energy loss models both with and without coalescence.}
\end{minipage} 
\end{figure}

In Fig.\ \ref{fig:raa_coal} we compare the $R_{AA}$ for PbPb collisions at 5.02 TeV for our two different ``best fit" models: the Langevin description and energy loss (see \cite{Katz:2019fkc} for further comparisons of energy loss fluctuations, initial state models, diffusion coefficients, and energy loss parameterizations).  Furthermore, we compare both models with and without coalescence. It is clear that at low $p_T$ the need for coalescence is much more obvious whereas above $p_T> 5 $ GeV, the need for coalescence is minimal.  

When it comes to $v_2\left\{2\right\}$ the influence of coalescence appears to play a role even in the energy loss results up to roughly $p_T\sim 10$ GeV.  For both Langevin and energy loss the effect of coalescence shifts the peak in the $v_2$ curve to the right (higher $p_T$).  Additionally, the $v_2$ results make it quite clear that energy loss is needed above $p_T> 5 $ GeV because the Langevin results significantly under-predict the data whereas the energy loss model can reproduce the experimental data perfectly.

\begin{figure}[h]
\centering
\includegraphics[width=\linewidth]{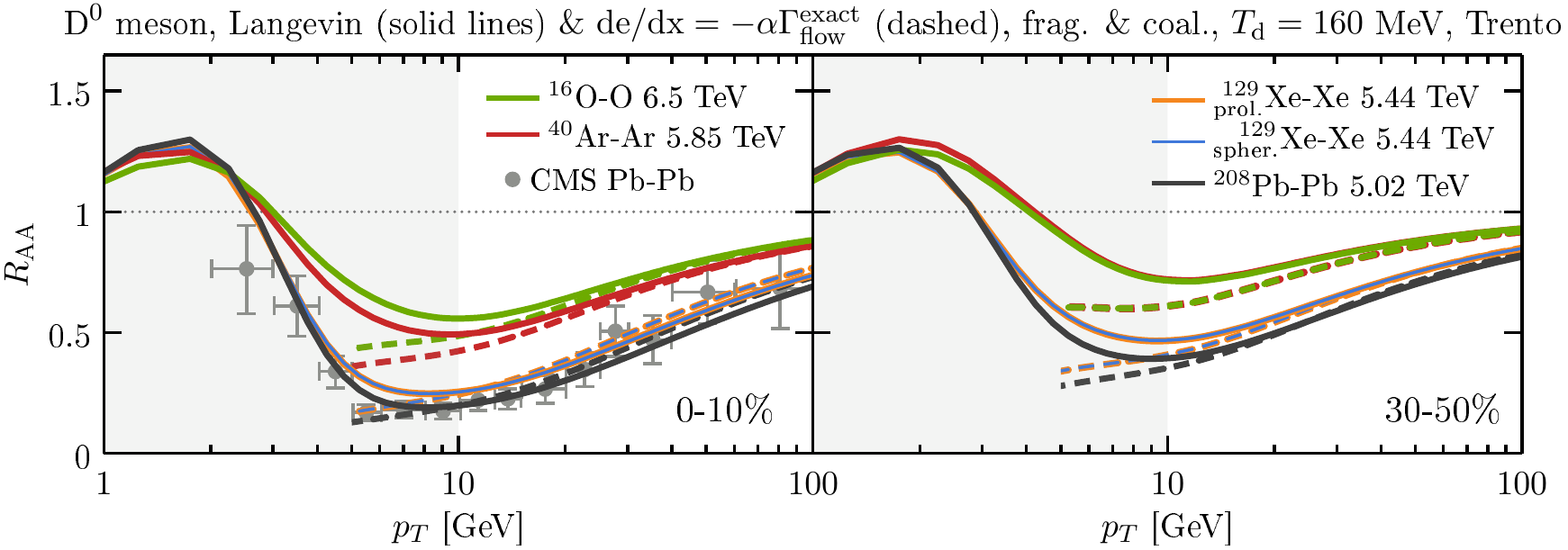}
\caption{\label{fig:raa} Plot of $R_{AA}$ across different collisional systems (i.e. system sizes), comparing two different centrality classes: $0-10\%$ and $30-50\%$.}
\end{figure}

Using then our best fit models (always including coalescence) we then make predictions across system size in Fig. \ref{fig:raa} for the nuclear modification factor.  We find that the smaller systems, indeed, bring $\raa$ closer to 1 but still have rather significant deviations from 1 at $p_T\sim 10$ GeV where a minimum is seen.  Generally $30-50\%$ centrality class can get an $\raa \sim 0.8$ at its minimum whereas  $0-10\%$ centrality class has a minimum closer to 0.5 for our smallest system size of OO collisions. We note that OO collisions are about double the radius of pPb collisions. Finally, we see that Langevin and energy loss descriptions are nearly identical in $\raa$ at high $p_T$ but between $p_T\sim 5-10$ GeV the energy loss model predicts significantly more heavy flavor suppression than the Langevin model.  This point specifically is interesting because it may be that in smaller systems a Langevin description, which predicts less suppression, is preferred in contrast to large systems where Langevin matches data only at low $p_T$.

\begin{figure}[h]
\centering
\includegraphics[width=0.5\linewidth]{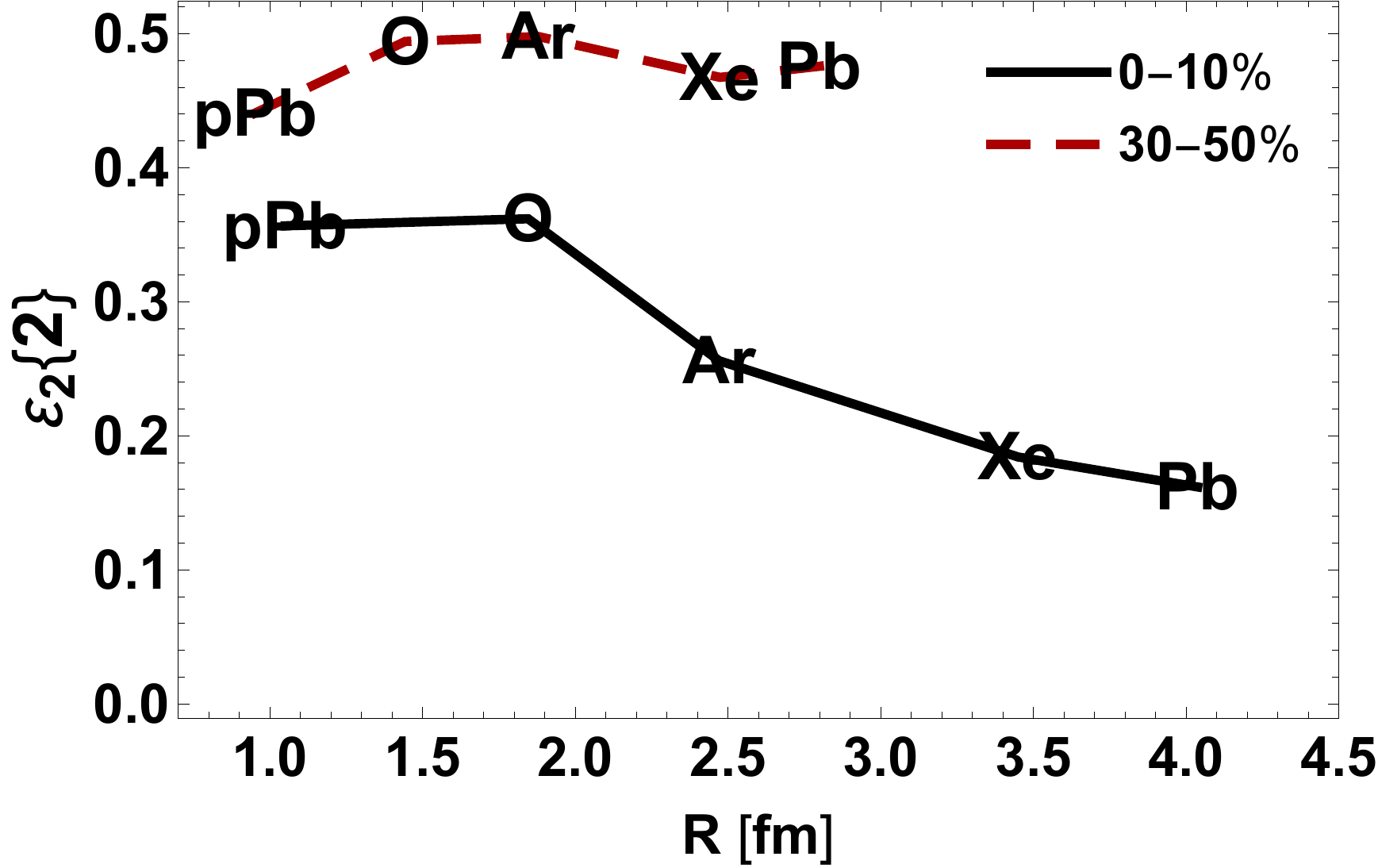}
\caption{\label{fig:ecc} Plot of elliptical eccentricities versus average radius for different collisional systems, comparing two different centrality classes: $0-10\%$ and $30-50\%$.}
\end{figure}

Because we study $v_2$, it is important to study the shape of the initial state through the eccentricities.  It has been shown that the initial eccentricities are very strongly correlated with the final $v_2$ in both the soft \cite{Teaney:2010vd,Gardim:2011xv,Niemi:2012aj,Teaney:2012ke,Qiu:2011iv,Gardim:2014tya} and the hard/heavy sector \cite{betz:2016ayq,noronha-hostler:2016eow, prado:2016szr, sirunyan:2017pan,Andres:2019eus,Katz:2019fkc,Katz:2019qwv}. Thus, studying the eccentricities across system size can shed light on the anticipated final $v_2$ of D mesons. In Fig.\ \ref{fig:ecc} we perform precisely this comparison and find quite different results depending on the centrality class.  In centrality collisions of $0-10\%$ there is a strong system size dependence in $\varepsilon_2$ such that as the system size decreases, the eccentricities increase.  In contrast, in midcentral collisions of $30-50\%$ we find that the eccentricities are nearly constant and only the system size changes.  Thus, $30-50\%$ is the best centrality class to actually observe system size effects (to reduce the competing effect of variation in eccentricities).

\begin{figure}[h]
\centering
\includegraphics[width=\linewidth]{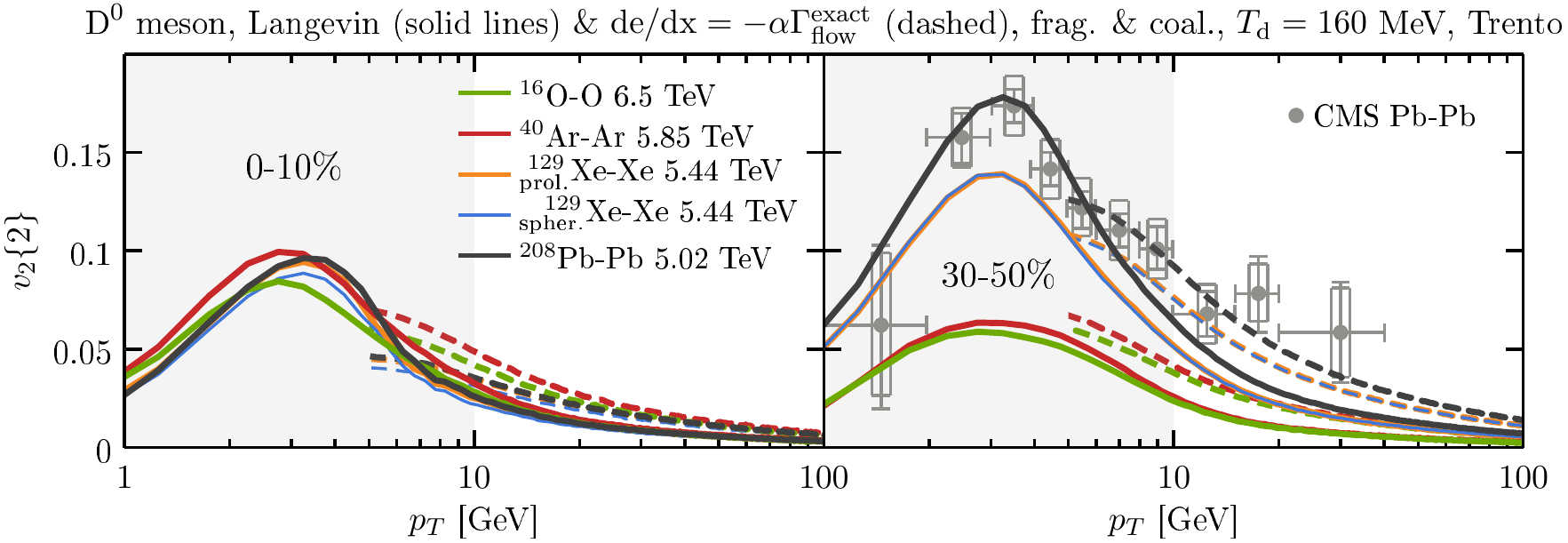}
\caption{\label{fig:v2} Plot of $v_2\left\{2\right\}$ across different collisional systems (i.e. system sizes), comparing two different centrality classes: $0-10\%$ and $30-50\%$.}
\end{figure}

Then in Fig.\ \ref{fig:v2} we make the full comparison of the $v_2$ calculations.  We find that as anticipated from the eccentricities that $0-10\%$ paints a complicated picture where there is very little system size dependence.  This is because the increase in $\varepsilon_2$ in small systems would have the effect of increasing $v_2$ whereas the system size decrease (demonstrated through the average radius, $R$) would suppress $v_2$.  Overall, these two competing effects lead to a nearly identical $v_2$ across system size.  In contrast, the $30-50\%$ centrality class sees a very clear suppression of $v_2$ as one decrease the system size.  

Additionally, in Fig.\ \ref{fig:v2} we see that a prolate nucleus leads to a larger $v_2$ for D mesons.  This is somewhat surprising because typically the effect of a deformation only shows up in very central collisions.  Additionally, one would likely expect that D mesons are less sensitive to such small deformation, however, it appears that they indeed can see these differences. 

\section{Conclusions}

Similar to what was found in the soft sector in \cite{Sievert:2019zjr}, different behaviors can appear when scaling either by multiplicity (see also \cite{Acharya:2019vdf}) or centrality.  In this talk, we argue that when scaling by centrality the $v_2$ results have a complex interplay between eccentricities and shrinking system size such that the D meson $v_2$ is nearly identical across system size in most central collisions.  However, we argue that comparing $30-50\%$ centrality allows a focus  only on system size effects while keeping the eccentricities nearly identical.  In this case we predict that D meson $v_2$ will be significantly suppressed in smaller systems in $30-50\%$ centrality, this result occurs both in Langevin and energy loss descriptions.

One surprising feature is that we do find a sensitive of D meson $v_2$ to a deformed  $^{129}\Xe$ nucleus in $0-10\%$ collisions in the range of $p_T=3-5$ GeV, in light of these results it would be quite interesting to revisit RHIC UU results in central collisions in order to study D meson $v_2$ or to also explore effects of deformed nuclei in the future sPHENIX experiment.  

Finally, we study the system size dependence of a heavy flavor Langevin description versus energy loss description.  While the overage magnitude shifts somewhat between the two descriptions, we find that they are both influenced by system size in roughly equivalent manners. In the case of $\raa$ they both approach 1 as the system size is decreased (however, a Langevin description is slightly closer to 1 compared to the energy loss description). However, both still are quite a bit farther away from the $R_{pPb}\sim 1$ result.  The $v_2$ results for energy loss tend to produce more $v_2$ for $p_T>5$ GeV (bringing it closer to experimental data in  PbPb collisions than the Langevin models at high $p_T$).  In future work we hope to extent our model to even smaller systems such as pPb and pp to compare to recent results from ATLAS and CMS. Additionally, we expect a similar effect to occur at RHIC (as discussed in \cite{Huang:2019tgz}). Furthermore, it will be interesting to explore further soft-heavy correlations such as \cite{Plumari:2019yhg} or study the initialization time of heavy flavor dynamics more carefully \cite{Andres:2019eus}. Furthermore, it would be interesting to explore these ideas further in proposed measurements such as polarized beams~\cite{Bozek:2018xzy} and ultracentral deformed ion-ion collisions~\cite{Noronha-Hostler:2019ytn}.

\section{Acknowledgments.} The authors thank Funda\c{c}\~ao de Amparo \`a Pesquisa do Estado de S\~ao Paulo (FAPESP) and Conselho Nacional de Desenvolvimento Cient\'ifico e Tecnol\'ogico (CNPq) for support. R.K. is supported by the Region Pays de la Loire (France) under contract No. 2015-08473. C.A.G.P. is supported by the NSFC under grant No. 11521064, MOST of China under Project No. 2014CB845404.  J.N.H. acknowledges the support of the Alfred P. Sloan Foundation; support from the US-DOE Nuclear Science Grant No. DE-SC0019175; the Illinois Campus Cluster, a computing resource that is operated by the Illinois Campus Cluster Program (ICCP) in conjunction with the National Center for Supercomputing Applications (NCSA) and which is supported by funds from the University of Illinois at Urbana-Champaign; and the Office of Advanced Research Computing (OARC) at Rutgers, The State University of New Jersey for providing access to the Amarel cluster and associated research computing resources that have contributed to the results reported here.

\section*{References}
\bibliography{BIG}

\providecommand{\newblock}{}
\begin{thebibliography}{10}
\expandafter\ifx\csname url\endcsname\relax
  \def\url#1{{\tt #1}}\fi
\expandafter\ifx\csname urlprefix\endcsname\relax\def\urlprefix{URL }\fi
\providecommand{\eprint}[2][]{\url{#2}}

\bibitem{Song:2010mg}
Song H, Bass S~A, Heinz U, Hirano T and Shen C 2011 {\em Phys. Rev. Lett.\/}
  {\bf 106} 192301 [Erratum: Phys.Rev.Lett. 109, 139904 (2012)]
  (\textit{Preprint} \eprint{1011.2783})

\bibitem{Bozek:2012qs}
Bozek P and Wyskiel-Piekarska I 2012 {\em Phys. Rev. C\/} {\bf 85} 064915
  (\textit{Preprint} \eprint{1203.6513})

\bibitem{Gardim:2012yp}
Gardim F~G, Grassi F, Luzum M and Ollitrault J~Y 2012 {\em Phys. Rev. Lett.\/}
  {\bf 109} 202302 (\textit{Preprint} \eprint{1203.2882})

\bibitem{Bozek:2013uha}
Bozek P and Broniowski W 2013 {\em Phys. Rev.\/} {\bf C88} 014903
  (\textit{Preprint} \eprint{1304.3044})

\bibitem{Niemi:2015qia}
Niemi H, Eskola K~J and Paatelainen R 2016 {\em Phys. Rev.\/} {\bf C93} 024907
  (\textit{Preprint} \eprint{1505.02677})

\bibitem{Ryu:2015vwa}
Ryu S, Paquet J~F, Shen C, Denicol G~S, Schenke B, Jeon S and Gale C 2015 {\em
  Phys. Rev. Lett.\/} {\bf 115} 132301 (\textit{Preprint} \eprint{1502.01675})

\bibitem{McDonald:2016vlt}
McDonald S, Shen C, Fillion-Gourdeau F, Jeon S and Gale C 2017 {\em Phys.
  Rev.\/} {\bf C95} 064913 (\textit{Preprint} \eprint{1609.02958})

\bibitem{Bernhard:2016tnd}
Bernhard J~E, Moreland J~S, Bass S~A, Liu J and Heinz U 2016 {\em Phys. Rev.\/}
  {\bf C94} 024907 (\textit{Preprint} \eprint{1605.03954})

\bibitem{Gardim:2016nrr}
Gardim F~G, Grassi F, Luzum M and Noronha-Hostler J 2017 {\em Phys. Rev.\/}
  {\bf C95} 034901 (\textit{Preprint} \eprint{1608.02982})

\bibitem{Giacalone:2016afq}
Giacalone G, Yan L, Noronha-Hostler J and Ollitrault J~Y 2016 {\em Phys. Rev.
  C\/} {\bf 94} 014906 (\textit{Preprint} \eprint{1605.08303})

\bibitem{Alba:2017hhe}
Alba P, Mantovani~Sarti V, Noronha J, Noronha-Hostler J, Parotto P,
  Portillo~Vazquez I and Ratti C 2018 {\em Phys. Rev.\/} {\bf C98} 034909
  (\textit{Preprint} \eprint{1711.05207})

\bibitem{Gardim:2017ruc}
Gardim F~G, Grassi F, Ishida P, Luzum M, Magalhães P~S and Noronha-Hostler J
  2018 {\em Phys. Rev.\/} {\bf C97} 064919 (\textit{Preprint}
  \eprint{1712.03912})

\bibitem{Giacalone:2017dud}
Giacalone G, Noronha-Hostler J, Luzum M and Ollitrault J~Y 2018 {\em Phys.
  Rev.\/} {\bf C97} 034904 (\textit{Preprint} \eprint{1711.08499})

\bibitem{Eskola:2017bup}
Eskola K~J, Niemi H, Paatelainen R and Tuominen K 2018 {\em Phys. Rev.\/} {\bf
  C97} 034911 (\textit{Preprint} \eprint{1711.09803})

\bibitem{Weller:2017tsr}
Weller R~D and Romatschke P 2017 {\em Phys. Lett.\/} {\bf B774} 351--356
  (\textit{Preprint} \eprint{1701.07145})

\bibitem{Schenke:2019ruo}
Schenke B, Shen C and Tribedy P 2019  (\textit{Preprint} \eprint{1901.04378})

\bibitem{Giacalone:2020lbm}
Giacalone G, Gardim F~G, Noronha-Hostler J and Ollitrault J~Y 2020
  (\textit{Preprint} \eprint{2004.09799})

\bibitem{Giacalone:2020dln}
Giacalone G, Gardim F~G, Noronha-Hostler J and Ollitrault J~Y 2020
  (\textit{Preprint} \eprint{2004.01765})

\bibitem{Nahrgang:2014vza}
Nahrgang M, Aichelin J, Bass S, Gossiaux P~B and Werner K 2015 {\em Phys.
  Rev.\/} {\bf C91} 014904 (\textit{Preprint} \eprint{1410.5396})

\bibitem{betz:2016ayq}
Betz B, Gyulassy M, Luzum M, Noronha J, Noronha-Hostler J, Portillo I and Ratti
  C 2017 {\em Phys. Rev.\/} {\bf C95} 044901 (\textit{Preprint}
  \eprint{1609.05171})

\bibitem{noronha-hostler:2016eow}
Noronha-Hostler J, Betz B, Noronha J and Gyulassy M 2016 {\em Phys. Rev.
  Lett.\/} {\bf 116} 252301 (\textit{Preprint} \eprint{1602.03788})

\bibitem{prado:2016szr}
Prado C~A~G, Noronha-Hostler J, Katz R, Suaide A~A~P, Noronha J, Munhoz M~G and
  Cosentino M~R 2017 {\em Phys. Rev.\/} {\bf C96} 064903 (\textit{Preprint}
  \eprint{1611.02965})

\bibitem{sirunyan:2017pan}
Sirunyan A~M {\em et~al.\/} (CMS) 2018 {\em Phys. Lett.\/} {\bf B776} 195--216
  (\textit{Preprint} \eprint{1702.00630})

\bibitem{Andres:2019eus}
Andres C, Armesto N, Niemi H, Paatelainen R and Salgado C~A 2019
  (\textit{Preprint} \eprint{1902.03231})

\bibitem{Katz:2019qwv}
Katz R, Prado C~A, Noronha-Hostler J and Suaide A~A 2019  (\textit{Preprint}
  \eprint{1907.03308})

\bibitem{Chatrchyan:2013nka}
Chatrchyan S {\em et~al.\/} (CMS) 2013 {\em Phys. Lett.\/} {\bf B724} 213--240
  (\textit{Preprint} \eprint{1305.0609})

\bibitem{Aaboud:2017acw}
Aaboud M {\em et~al.\/} (ATLAS) 2017 {\em Eur. Phys. J.\/} {\bf C77} 428
  (\textit{Preprint} \eprint{1705.04176})

\bibitem{Aaboud:2017blb}
Aaboud M {\em et~al.\/} (ATLAS) 2018 {\em Phys. Rev.\/} {\bf C97} 024904
  (\textit{Preprint} \eprint{1708.03559})

\bibitem{Aad:2013fja}
Aad G {\em et~al.\/} (ATLAS) 2013 {\em Phys. Lett.\/} {\bf B725} 60--78
  (\textit{Preprint} \eprint{1303.2084})

\bibitem{sirunyan:2018toe}
Sirunyan A~M {\em et~al.\/} (CMS) 2018 {\em Phys. Rev. Lett.\/} {\bf 121}
  082301 (\textit{Preprint} \eprint{1804.09767})

\bibitem{Khachatryan:2014jra}
Khachatryan V {\em et~al.\/} (CMS) 2015 {\em Phys. Lett.\/} {\bf B742} 200--224
  (\textit{Preprint} \eprint{1409.3392})

\bibitem{Khachatryan:2015waa}
Khachatryan V {\em et~al.\/} (CMS) 2015 {\em Phys. Rev. Lett.\/} {\bf 115}
  012301 (\textit{Preprint} \eprint{1502.05382})

\bibitem{Khachatryan:2015oea}
Khachatryan V {\em et~al.\/} (CMS) 2015 {\em Phys. Rev.\/} {\bf C92} 034911
  (\textit{Preprint} \eprint{1503.01692})

\bibitem{Sirunyan:2017uyl}
Sirunyan A~M {\em et~al.\/} (CMS) 2018 {\em Phys. Rev. Lett.\/} {\bf 120}
  092301 (\textit{Preprint} \eprint{1709.09189})

\bibitem{ABELEV:2013wsa}
Abelev B~B {\em et~al.\/} (ALICE) 2013 {\em Phys. Lett.\/} {\bf B726} 164--177
  (\textit{Preprint} \eprint{1307.3237})

\bibitem{Abelev:2014mda}
Abelev B~B {\em et~al.\/} (ALICE) 2014 {\em Phys. Rev.\/} {\bf C90} 054901
  (\textit{Preprint} \eprint{1406.2474})

\bibitem{Adare:2013piz}
Adare A {\em et~al.\/} (PHENIX) 2013 {\em Phys. Rev. Lett.\/} {\bf 111} 212301
  (\textit{Preprint} \eprint{1303.1794})

\bibitem{Adare:2014keg}
Adare A {\em et~al.\/} (PHENIX) 2015 {\em Phys. Rev. Lett.\/} {\bf 114} 192301
  (\textit{Preprint} \eprint{1404.7461})

\bibitem{Aidala:2018mcw}
Aidala C {\em et~al.\/} (PHENIX) 2018  (\textit{Preprint} \eprint{1805.02973})

\bibitem{Adare:2018toe}
Adare A {\em et~al.\/} (PHENIX) 2018  (\textit{Preprint} \eprint{1807.11928})

\bibitem{Adare:2015ctn}
Adare A {\em et~al.\/} (PHENIX) 2015 {\em Phys. Rev. Lett.\/} {\bf 115} 142301
  (\textit{Preprint} \eprint{1507.06273})

\bibitem{Aidala:2016vgl}
Aidala C {\em et~al.\/} 2017 {\em Phys. Rev.\/} {\bf C95} 034910
  (\textit{Preprint} \eprint{1609.02894})

\bibitem{Adare:2017wlc}
Adare A {\em et~al.\/} (PHENIX) 2018 {\em Phys. Rev.\/} {\bf C97} 064904
  (\textit{Preprint} \eprint{1710.09736})

\bibitem{Adare:2017rdq}
Adare A {\em et~al.\/} (PHENIX) 2018 {\em Phys. Rev.\/} {\bf C98} 014912
  (\textit{Preprint} \eprint{1711.09003})

\bibitem{Aidala:2017pup}
Aidala C {\em et~al.\/} (PHENIX) 2017 {\em Phys. Rev.\/} {\bf C96} 064905
  (\textit{Preprint} \eprint{1708.06983})

\bibitem{Aidala:2017ajz}
Aidala C {\em et~al.\/} (PHENIX) 2018 {\em Phys. Rev. Lett.\/} {\bf 120} 062302
  (\textit{Preprint} \eprint{1707.06108})

\bibitem{ALICE:2017jyt}
Adam J {\em et~al.\/} (ALICE) 2017 {\em Nature Phys.\/} {\bf 13} 535--539
  (\textit{Preprint} \eprint{1606.07424})

\bibitem{Bozek:2011if}
Bozek P 2012 {\em Phys. Rev.\/} {\bf C85} 014911 (\textit{Preprint}
  \eprint{1112.0915})

\bibitem{Bozek:2012gr}
Bozek P and Broniowski W 2013 {\em Phys. Lett.\/} {\bf B718} 1557--1561
  (\textit{Preprint} \eprint{1211.0845})

\bibitem{Bozek:2013ska}
Bozek P, Broniowski W and Torrieri G 2013 {\em Phys. Rev. Lett.\/} {\bf 111}
  172303 (\textit{Preprint} \eprint{1307.5060})

\bibitem{Kozlov:2014fqa}
Kozlov I, Luzum M, Denicol G, Jeon S and Gale C 2014  (\textit{Preprint}
  \eprint{1405.3976})

\bibitem{Zhou:2015iba}
Zhou Y, Zhu X, Li P and Song H 2015 {\em Phys. Rev.\/} {\bf C91} 064908
  (\textit{Preprint} \eprint{1503.06986})

\bibitem{Zhao:2017rgg}
Zhao W, Zhou Y, Xu H, Deng W and Song H 2018 {\em Phys. Lett.\/} {\bf B780}
  495--500 (\textit{Preprint} \eprint{1801.00271})

\bibitem{Mantysaari:2017cni}
Mäntysaari H, Schenke B, Shen C and Tribedy P 2017 {\em Phys. Lett.\/} {\bf
  B772} 681--686 (\textit{Preprint} \eprint{1705.03177})

\bibitem{Adam:2015qda}
Adam J {\em et~al.\/} (ALICE) 2016 {\em Phys. Lett. B\/} {\bf 754} 81--93
  (\textit{Preprint} \eprint{1509.07491})

\bibitem{Acharya:2019mno}
Acharya S {\em et~al.\/} (ALICE) 2019 {\em JHEP\/} {\bf 12} 092
  (\textit{Preprint} \eprint{1906.03425})

\bibitem{Perepelitsa:2020pcf}
Perepelitsa D~V 2020 (\textit{Preprint} \eprint{2005.05981})

\bibitem{Zhang:2019dth}
Zhang C, Marquet C, Qin G~Y, Wei S~Y and Xiao B~W 2019 {\em Phys. Rev. Lett.\/}
  {\bf 122} 172302 (\textit{Preprint} \eprint{1901.10320})

\bibitem{Xu:2015iha}
Xu Y, Cao S, Qin G~Y, Ke W, Nahrgang M, Auvinen J and Bass S~A 2016 {\em Nucl.
  Part. Phys. Proc.\/} {\bf 276-278} 225--228 (\textit{Preprint}
  \eprint{1510.07520})

\bibitem{Kang:2014hha}
Kang Z~B, Vitev I, Wang E, Xing H and Zhang C 2015 {\em Phys. Lett.\/} {\bf
  B740} 23--29 (\textit{Preprint} \eprint{1409.2494})

\bibitem{Sharma:2009hn}
Sharma R, Vitev I and Zhang B~W 2009 {\em Phys. Rev.\/} {\bf C80} 054902
  (\textit{Preprint} \eprint{0904.0032})

\bibitem{Citron:2018lsq}
Citron Z {\em et~al.\/} 2018 (\textit{Preprint} \eprint{1812.06772})

\bibitem{Katz:2019fkc}
Katz R, Prado C~A~G, Noronha-Hostler J, Noronha J and Suaide A~A~P 2019
  (\textit{Preprint} \eprint{1906.10768})

\bibitem{Cacciari:1998it}
Cacciari M, Greco M and Nason P 1998 {\em JHEP\/} {\bf 05} 007
  (\textit{Preprint} \eprint{hep-ph/9803400})

\bibitem{Cacciari:2001td}
Cacciari M, Frixione S and Nason P 2001 {\em JHEP\/} {\bf 03} 006
  (\textit{Preprint} \eprint{hep-ph/0102134})

\bibitem{Moreland:2014oya}
Moreland J~S, Bernhard J~E and Bass S~A 2015 {\em Phys. Rev.\/} {\bf C92}
  011901 (\textit{Preprint} \eprint{1412.4708})

\bibitem{Noronha-Hostler:2014dqa}
Noronha-Hostler J, Noronha J and Grassi F 2014 {\em Phys. Rev.\/} {\bf C90}
  034907 (\textit{Preprint} \eprint{1406.3333})

\bibitem{Noronha-Hostler:2013gga}
Noronha-Hostler J, Denicol G~S, Noronha J, Andrade R~P~G and Grassi F 2013 {\em
  Phys. Rev.\/} {\bf C88} 044916 (\textit{Preprint} \eprint{1305.1981})

\bibitem{Noronha-Hostler:2015coa}
Noronha-Hostler J, Noronha J and Gyulassy M 2016 {\em Phys. Rev. C\/} {\bf 93}
  024909 (\textit{Preprint} \eprint{1508.02455})

\bibitem{Moore:2004tg}
Moore G~D and Teaney D 2005 {\em Phys. Rev.\/} {\bf C71} 064904
  (\textit{Preprint} \eprint{hep-ph/0412346})

\bibitem{Aaboud:2018bdg}
Aaboud M {\em et~al.\/} (ATLAS) 2018 {\em Phys. Rev.\/} {\bf C98} 044905
  (\textit{Preprint} \eprint{1805.05220})

\bibitem{Giacalone:2017uqx}
Giacalone G, Noronha-Hostler J and Ollitrault J~Y 2017 {\em Phys. Rev.\/} {\bf
  C95} 054910 (\textit{Preprint} \eprint{1702.01730})

\bibitem{Rao:2019vgy}
Rao S, Sievert M and Noronha-Hostler J 2019  (\textit{Preprint}
  \eprint{1910.03677})

\bibitem{Sievert:2019zjr}
Sievert M~D and Noronha-Hostler J 2019  (\textit{Preprint} \eprint{1901.01319})

\bibitem{Borsanyi:2013bia}
Borsanyi S, Fodor Z, Hoelbling C, Katz S~D, Krieg S and Szabo K~K 2014 {\em
  Phys. Lett.\/} {\bf B730} 99--104 (\textit{Preprint} \eprint{1309.5258})

\bibitem{alba:2017mqu}
Alba P {\em et~al.\/} 2017 {\em Phys. Rev.\/} {\bf D96} 034517
  (\textit{Preprint} \eprint{1702.01113})

\bibitem{Alba:2020jir}
Alba P, Sarti V~M, Noronha-Hostler J, Parotto P, Portillo-Vazquez I, Ratti C
  and Stafford J 2020  (\textit{Preprint} \eprint{2002.12395})

\bibitem{Acharya:2017qps}
Acharya S {\em et~al.\/} (ALICE) 2018 {\em Phys. Rev. Lett.\/} {\bf 120} 102301
  (\textit{Preprint} \eprint{1707.01005})

\bibitem{Xu:2018gux}
Xu Y {\em et~al.\/} 2019 {\em Phys. Rev.\/} {\bf C99} 014902 (\textit{Preprint}
  \eprint{1809.10734})

\bibitem{Teaney:2010vd}
Teaney D and Yan L 2011 {\em Phys. Rev.\/} {\bf C83} 064904 (\textit{Preprint}
  \eprint{1010.1876})

\bibitem{Gardim:2011xv}
Gardim F~G, Grassi F, Luzum M and Ollitrault J~Y 2012 {\em Phys. Rev.\/} {\bf
  C85} 024908 (\textit{Preprint} \eprint{1111.6538})

\bibitem{Niemi:2012aj}
Niemi H, Denicol G~S, Holopainen H and Huovinen P 2013 {\em Phys. Rev.\/} {\bf
  C87} 054901 (\textit{Preprint} \eprint{1212.1008})

\bibitem{Teaney:2012ke}
Teaney D and Yan L 2012 {\em Phys. Rev.\/} {\bf C86} 044908 (\textit{Preprint}
  \eprint{1206.1905})

\bibitem{Qiu:2011iv}
Qiu Z and Heinz U~W 2011 {\em Phys. Rev.\/} {\bf C84} 024911 (\textit{Preprint}
  \eprint{1104.0650})

\bibitem{Gardim:2014tya}
Gardim F~G, Noronha-Hostler J, Luzum M and Grassi F 2015 {\em Phys. Rev.\/}
  {\bf C91} 034902 (\textit{Preprint} \eprint{1411.2574})

\bibitem{Acharya:2019vdf}
Acharya S {\em et~al.\/} (ALICE) 2019 {\em Phys. Rev. Lett.\/} {\bf 123} 142301
  (\textit{Preprint} \eprint{1903.01790})

\bibitem{Huang:2019tgz}
Huang S, Chen Z, Jia J and Li W 2019  (\textit{Preprint} \eprint{1904.10415})

\bibitem{Plumari:2019yhg}
Plumari S, Coci G, Das S~K, Minissale V and Greco V 2019 {\em Nucl. Phys.\/}
  {\bf A982} 655--658 (\textit{Preprint} \eprint{1901.07815})

\bibitem{Bozek:2018xzy}
Bozek P and Broniowski W 2018 {\em Phys. Rev. Lett.\/} {\bf 121} 202301
  (\textit{Preprint} \eprint{1808.09840})

\bibitem{Noronha-Hostler:2019ytn}
Noronha-Hostler J, Paladino N, Rao S, Sievert M~D and Wertepny D~E 2019
  (\textit{Preprint} \eprint{1905.13323})

\end{thebibliography}

\end{document}